\documentclass[11pt]{tibop-article}

\TIBauthor[O. Karras et al.]{
    Oliver Karras\affOne\orcid{0000-0001-5336-6899}\and
    Jan Göpfert\affTwo\affThree\orcid{0000-0003-4722-4012}\and
    Patrick Kuckertz\affTwo\orcid{0000-0002-2314-7107}\and
    Tristan Pelser\affTwo\affThree\orcid{0009-0006-0424-4784}\lastand Sören Auer\affOne\affFour\orcid{0000-0002-0698-2864}
}
\TIBaffiliations{
    \affOne TIB - Leibniz Information Centre for Science and Technology, Hanover, Germany\\
    \affTwo Forschungszentrum Jülich GmbH, Institute of Energy and Climate Research – Techno-economic Systems Analysis (IEK-3), Jülich, Germany\\
    \affThree RWTH Aachen University, Chair for Fuel Cells, Faculty of Mechanical Engineering, Aachen, Germany\\
    \affFour Leibniz University Hannover, L3S Research Center, Hannover, Germany
}

\TIBtitle{Organizing Scientific Knowledge From Energy System Research Using the Open Research Knowledge Graph}
\TIBbundlename[Organizing Scientific Knowledge From Energy System Research Using the ORKG]{NFDI4Energy Conference}
\TIBconferencesession{1. NFDI4Energy Conference}%ignored for journal articles
\TIBdoi{DOI: \url{https://zenodo.org/doi/10.5281/zenodo.10560077}}
\TIBkeywords{Energy System Research, FAIR, Open Science, Scenario, Scientific Knowledge, Scientific Communication, Infrastructure, Knowledge Graph, Ontology}
\usepackage{xfrac}
\usepackage[utf8]{inputenc}
\usepackage{listings}

\begin{document}
\maketitle

\section{Introduction}
%Words: 233
Engineering sciences, such as energy system research, play an important role in developing solutions to technical, environmental, economic, and social challenges of our modern society~\cite{Schmitt.2020, Niesse.2022}. In this context, the transformation of energy systems into climate-neutral systems is one of the key strategies for mitigating climate change. For the transformation of energy systems, engineers model, simulate and analyze scenarios and transformation pathways to initiate debates about possible transformation strategies. For these debates and research in general, all steps of the research process must be traceable to guarantee the trustworthiness of published results, avoid redundancies, and ensure their social acceptance~\cite{Schmitt.2020, Pfenninger.2017}. However, the analysis of energy systems is an interdisciplinary field as the investigations of large, complex energy systems often require the use of different software applications and large amounts of heterogeneous data~\cite{Niesse.2022}. Engineers must therefore communicate, understand, and (re)use heterogeneous scientific knowledge and data~\cite{Niesse.2022, Booshehri.2021}. Although the importance of FAIR~\cite{Wilkinson.2016} scientific knowledge and data in the engineering sciences and energy system research is increasing, little research has been conducted on this topic~\cite{Schmitt.2020, Niesse.2022}. When it comes to publishing scientific knowledge and data from publications, software, and datasets (such as models, scenarios, and simulations) openly available and transparent, energy system research lags behind other research domains~\cite{Pfenninger.2017}. According to Schmitt et al.~\cite{Schmitt.2020} and Nieße et al.~\cite{Niesse.2022}, engineers need technical support in the form of infrastructures, services, and terminologies to improve communication, understanding, and (re)use of scientific knowledge and data.

\section{Background}
%Words: 151
In 2020, the consortium National Research Data Infrastructure for Engineering Sciences\footnote{\url{https://nfdi4ing.de/}} (NFDI4Ing) began developing and deploying infrastructures, i.a., the Open Research Knowledge Graph\footnote{\url{https://orkg.org/}} (ORKG), and services, i.a., the Terminology Service\footnote{\url{https://terminology.tib.eu/ts}} (TS), for engineers to organize FAIR scientific knowledge and data~\cite{Schmitt.2020}. At the same time, first valuable building blocks for FAIR energy system research have been developed, such as the Open Energy Platform\footnote{\url{https://openenergy-platform.org/}} (OEP) and the Open Energy Ontology\footnote{\url{https://openenergy-platform.org/ontology/}} (OEO)~\cite{Booshehri.2021}. The ORKG is a cross-domain research knowledge graph combining manual crowdsourcing and (semi-)automated approaches for the production, curation, and (re)use of FAIR scientific knowledge from publications, software, and datasets~\cite{Auer.2020, Karras.2021, Auer.2023a, Stocker.2023}. The TS is a cross-domain service that supports the discovery, provision, design, and curation of ontologies~\cite{Stroemert.2023}. Based on the work of the task area Ellen\footnote{\url{https://nfdi4ing.de/archetypes/ellen/}} in NFDI4Ing, the ORKG integrated the TS that in turn curates the OEO, so that engineers can describe and organize scientific knowledge and data as so-called ORKG \textit{contributions}.

\section{Contribution}
%Words: 160
In this proposal for a presentation, we outline how the consortium National Research Data Infrastructure for the Interdisciplinary Energy System Research\footnote{\url{https://nfdi4energy.uol.de/}} (NFDI4Energy) and thus the energy system research community can benefit from the ORKG by organizing scientific knowledge and data to improve its communication, understanding, and (re)use. In this way, we want to illustrate how NFDI4Energy can build on the results of NFDI4Ing to strengthen the collaboration of both consortia towards a joint NFDI. For this purpose, we organized scientific knowledge and data from scenarios of two exemplary use cases in the ORKG. In the first use case, we organized 14 scenario factsheets\footnote{\url{https://openenergy-platform.org/factsheets/scenarios/}} of the OEP regarding the assumptions, the data used, the associated study, and the results. In the second use case, we organized 25 scenarios from publications on greenhouse gas (GHG) reduction studies for Germany, which have been collected and compared regarding reported electricity supply and installed capacities for different energy sources and the respective scenario goal by Robinius et al.~\cite{Robinius.2020}.

%Words: 87
In both cases, we used ORKG \textit{templates} to facilitate the extraction of information and ensure consistent modeling. ORKG \textit{templates} specify the structure of ORKG \textit{contributions} similar to SHACL shapes~\cite{Knublauch.2017} and can use ontologies (with the help of the integrated TS) so that the semantic descriptions of scientific knowledge are consistent and comparable across all considered publications. For example, we developed ORKG \textit{templates} for the scenario goal\footnote{\url{https://orkg.org/template/R153118/}} or the electricity supply\footnote{\url{https://orkg.org/template/R152170/}} and used the OEO to ensure clear definitions and the logical interpretation of different types of sectors\footnote{\url{https://terminology.tib.eu/ts/ontologies/oeo/terms?iri=http\%3A\%2F\%2Fopenenergy-platform.org\%2Fontology\%2Foeo\%2FOEO_00000367&subtab=graph}}.

\begin{figure}[h!]
    \includegraphics[width=\linewidth]{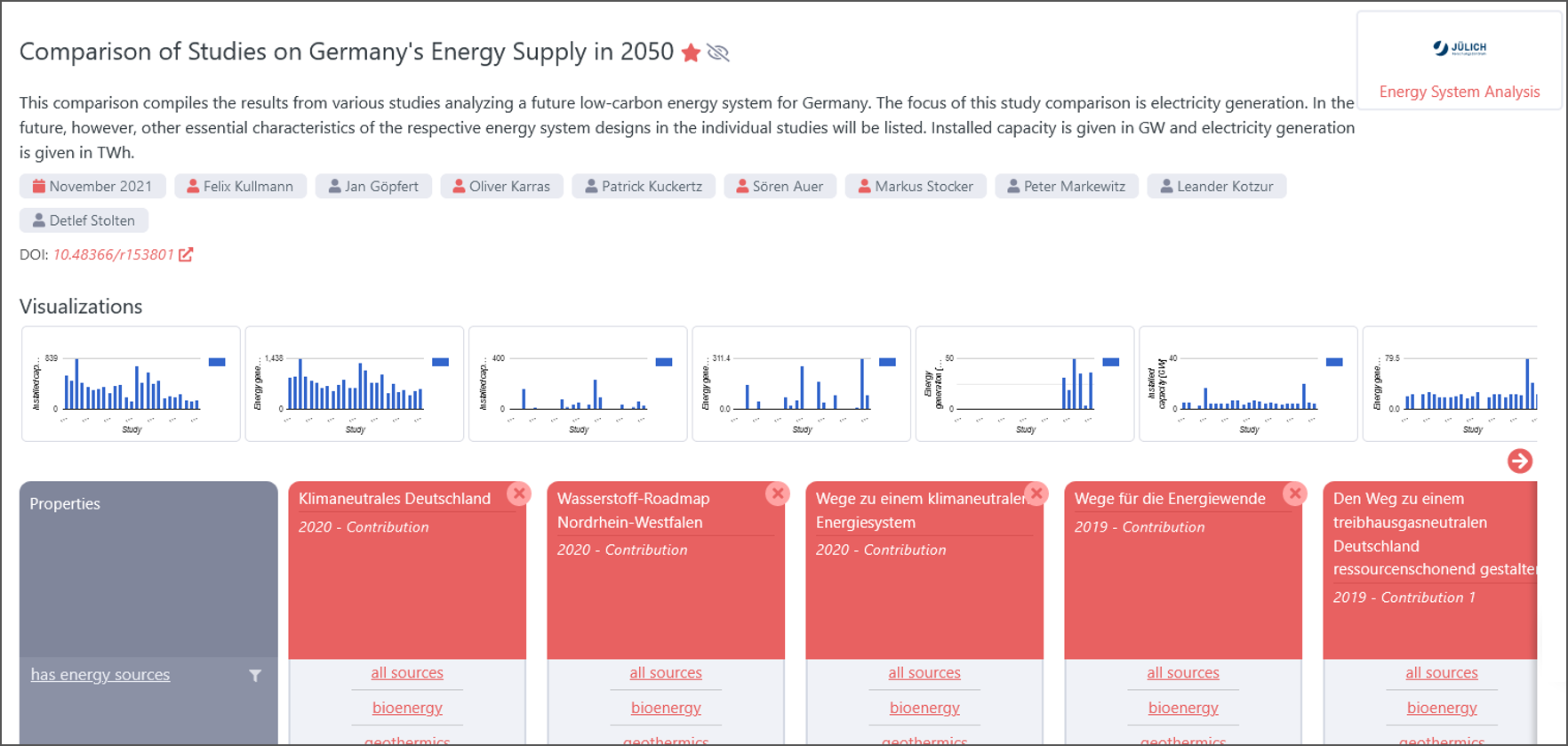}
    \caption{ORKG \textit{comparison} of 25 scenarios from GHG studies for Germany~\cite{Kullmann.2021}.}\label{fig:comparison}
\end{figure}

%Words: 164
For data analysis, we published an ORKG \textit{comparison} with a DOI for each use case to provide a referenceable, citable, and detailed overview of the scenario factsheets~\cite{Goepfert.2022} and GHG scenarios~\cite{Kullmann.2021} (see~\autoref{fig:comparison}). In contrast to the traditional way of publishing an overview of scenarios within a publication, ORKG \textit{comparisons} provide the benefit that they are versionable and can thus be continuously (re)used, updated, and expanded. When researchers publish new scenarios as factsheets or in publications, the ORKG \textit{comparisons} can be easily extended by describing the new scenarios using the same ORKG \textit{templates}, adding the new ORKG \textit{contribution} to the respective ORKG \textit{comparison}, and publishing the updated ORKG \textit{comparison} as a new version. The ORKG also supports the supplementation of ORKG \textit{comparisons} by creating visualizations based on the data contained therein either directly from the web frontend or via various access points, such as a REST API, a Python or R package, or a SPARQL endpoint, for example in combination with a Jupyter notebook.

\begin{figure}[h!]
    \includegraphics[width=.8\linewidth]{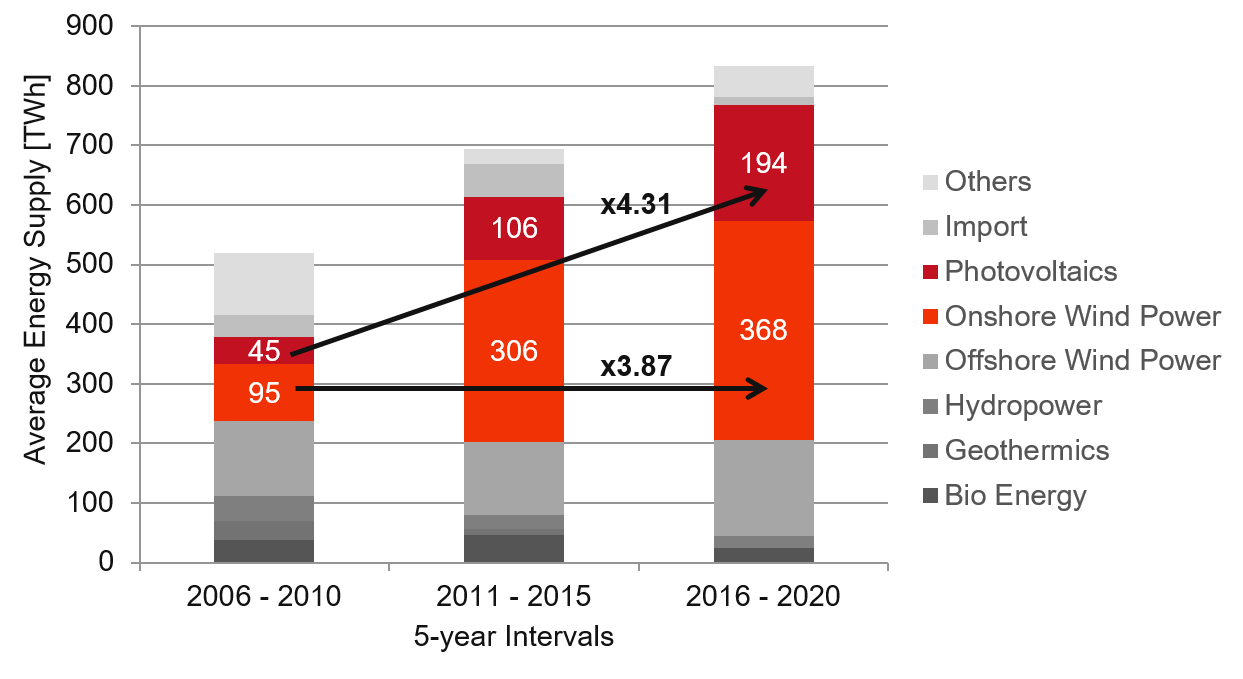}
    \caption{Visualized results from the SPARQL query by Auer et al.~\cite{Auer.2023}}\label{fig:results}
\end{figure}

%Words: 121
In addition, we established an ORKG \textit{observatory} on Energy System Research\footnote{\url{https://orkg.org/observatory/Energy_System_Research}}. The ORKG \textit{observatory} serves as a central access point to all related curated publications, comparisons, and visualizations so that other researchers can easily explore the content. For example, Auer et al.~\cite{Auer.2023} already reused the curated scientific knowledge from our two ORKG \textit{comparisons} by identifying and answering further natural language competency questions from domain experts beyond the previous consideration. For this purpose, they specified the competency question as SPARQL query (see \autoref{lst:query}). We executed this query on the SPARQL endpoint and visualized the results in \autoref{fig:results}. In particular, these results show that average energy supply from photovoltaics and onshore wind power increased approximately fourfold from the 2006 -- 2010 interval to the 2016 -- 2020 interval.

\newpage
\begin{center}
{\small
\begin{lstlisting}[language=SPARQL, label=lst:query]
SELECT ?range ?srcLabel AVG(?val) AS ?avgVal
WHERE {
  r:R153801  p:compareContribution  ?contrib.
  ?paper     p:hasContribution      ?contrib;
             p:hasPublicationYear   ?year.
  BIND(xsd:int(?year) AS ?y).
  VALUES(?range ?min ?max) {
    ("2001-2005" 2001 2005)
    ("2006-2010" 2006 2010)
    ("2011-2015" 2011 2015)
    ("2016-2020" 2016 2020)
  } FILTER(?min <= ?y && ?y <= ?max).
  ?contrib    p:hasEnergySources  ?energySrc.
  ?energySrc  rdfs:label          ?srcLabel;
              p:hasGeneration     ?energyGen.
  ?energyGen  p:hasValue          ?genVal.
  BIND(xsd:float(?genVal) AS ?val).
} ORDER BY ASC(?range)
\end{lstlisting}
}
\end{center}

\section{Conclusion}
%Words: 86
Overall, this proposal addresses multiple themes of the 1st NFDI4Energy conference\footnote{\url{https://nfdi4energy.uol.de/sites/conference/}}. In particular, we demonstrate how engineers can use the ORKG infrastructure in innovative ways to organize scientific knowledge and data from publications, software, and datasets by using ontologies, such as the OEO, for FAIR data management and thus open science. We hope to encourage and motivate engineers, especially in energy system research, to use the ORKG and its integrated services to communicate, understand, and (re)use scientific knowledge, thus promoting collaboration between NFDI4Energy and NFDI4Ing.

\section*{Data availability statement}
All data used are openly available in the Open Research Knowledge Graph: \url{https://orkg.org/} and in particular in the ORKG \textit{observatory} on Energy System Research: \url{https://orkg.org/observatory/Energy_System_Research}.

\section*{Author contributions}
\textbf{Oliver Karras}: Conceptualization, Methodology, Software, Validation, Formal analysis, Investigation, Resources, Data Curation, Writing - Original Draft, Visualization, Supervision, Project administration.\\
\textbf{Jan Göpfert}: Conceptualization, Methodology, Validation, Investigation, Resources, Data Curation, Writing - Original Draft.\\
\textbf{Patrick Kuckertz}: Conceptualization, Writing - Review \& Editing, Project administration.\\
\textbf{Tristan Pelser}: Writing - Review \& Editing.\\
\textbf{Sören Auer}: Conceptualization, Supervision, Funding acquisition.

\section*{Competing interests}
The authors declare that they have no competing interests.

\section*{Funding}
The authors thank the Federal Government, the Heads of Government of the Länder, as well as the Joint Science Conference (GWK), for their funding and support within the NFDI4Ing, NFDI4DataScience, and NFDI4Energy consortia. This work was funded by the German Research Foundation (DFG) -project numbers 442146713, 460234259, 501865131, by the European Research Council for the project ScienceGRAPH (Grant agreement ID: 819536), and by the TIB - Leibniz Information Centre for Science and Technology.

\section*{Acknowledgements}
The ORKG is a team effort comprising overall more than 50 people distributed worldwide working on the development of the infrastructure and its services, curating content, and researching novel features.

\printbibliography[heading=references]

@misc{Schmitt.2020,
  author       = {Schmitt, Robert H. and
                  Anthofer, Verena and
                  Auer, Sören and
                  Başkaya, Sait and
                  Bischof, Christian and
                  Bronger, Torsten and
                  Claus, Florian and
                  Cordes, Florian and
                  Demandt, Évariste and
                  Eifert, Thomas and
                  Flemisch, Bernd and
                  Fuchs, Matthias and
                  Fuhrmans, Marc and
                  Gerike, Regine and
                  Gerstner, Eva-Maria and
                  Hanke, Vanessa and
                  Heine, Ina and
                  Huebser, Louis and
                  Iglezakis, Dorothea and
                  Jagusch, Gerald and
                  Klinger, Axel and
                  Krafczyk, Manfred and
                  Kraft, Angelina and
                  Kuckertz, Patrick and
                  Küsters, Ulrike and
                  Lachmayer, Roland and
                  Langenbach, Christian and
                  Mozgova, Iryna and
                  Müller, Matthias S. and
                  Nestler, Britta and
                  Pelz, Peter and
                  Politze, Marius and
                  Preuß, Nils and
                  Przybylski-Freund, Marie-Dominique and
                  Rißler-Pipka, Nanette and
                  Robinius, Martin and
                  Schachtner, Joachim and
                  Schlenz, Hartmut and
                  Schwarz, Annett and
                  Schwibs, Jürgen and
                  Selzer, Michael and
                  Sens, Irina and
                  Stäcker, Thomas and
                  Stemmer, Christian and
                  Stille, Wolfgang and
                  Stolten, Detlef and
                  Stotzka, Rainer and
                  Streit, Achim and
                  Strötgen, Robert and
                  Wang, Wei Min},
  title        = {{NFDI4Ing - The National Research Data 
                   Infrastructure for Engineering Sciences}},
  year         = 2020,
  publisher    = {Zenodo},
  doi          = {10.5281/zenodo.4015201}
}

@misc{Niesse.2022,
  author       = {Nieße, Astrid and
                  Ferenz, Stephan and
                  Auer, Sören and
                  Dähling, Stefan and
                  Decker, Stefan and
                  Dorfner, Johannes and
                  German, Reinhard and
                  Gütlein, Moritz and
                  Hagenmeyer, Veit and
                  Henni, Sarah and
                  Lehnhoff, Sebastian and
                  Lilliestam, Johan and
                  Lu, Linna and
                  Mey, Franziska and
                  Monti, Antonello and
                  Muschner, Christoph and
                  Reinkensmeier, Jan and
                  Richter, Mascha and
                  Schäfer, Mirko and
                  Staudt, Philipp and
                  Süß, Wolfgang and
                  Vogel, Berthold and
                  Weinhardt, Christof and
                  Weidlich, Anke and
                  Zilles, Julia},
  title        = {{NFDI4Energy – National Research Data 
                   Infrastructure for the Interdisciplinary Energy
                   System Research}},
  year         = 2022,
  publisher    = {Zenodo},
  doi          = {10.5281/zenodo.6772013},
}

@article{Pfenninger.2017,
title = {{The Importance of Open Data and Software: Is Energy Research Lagging Behind?}},
journal = {Energy Policy},
volume = {101},
pages = {211-215},
year = {2017},
doi = {10.1016/j.enpol.2016.11.046},
author = {Stefan Pfenninger and Joseph DeCarolis and Lion Hirth and Sylvain Quoilin and Iain Staffell}
}

@article{Booshehri.2021,
title = {{Introducing the Open Energy Ontology: Enhancing Data Interpretation and Interfacing in Energy Systems Analysis}},
journal = {Energy and AI},
volume = {5},
year = {2021},
doi = {10.1016/j.egyai.2021.100074},
author = {Meisam Booshehri and Lukas Emele and Simon Flügel and Hannah Förster and Johannes Frey and Ulrich Frey and Martin Glauer and Janna Hastings and Christian Hofmann and Carsten Hoyer-Klick and Ludwig Hülk and Anna Kleinau and Kevin Knosala and Leander Kotzur and Patrick Kuckertz and Till Mossakowski and Christoph Muschner and Fabian Neuhaus and Michaja Pehl and Martin Robinius and Vera Sehn and Mirjam Stappel}
}

@article{Wilkinson.2016,
  title={{The FAIR Guiding Principles for Scientific Data Management and Stewardship}},
  author={Wilkinson, Mark D and Dumontier, Michel and Aalbersberg, IJsbrand Jan and Appleton, Gabrielle and Axton, Myles and others},
  journal={Scientific Data},
  volume={3},
  number={1},
  year={2016},
  publisher={Nature Publishing Group},
  doi={10.1038/sdata.2016.18}
}

@article{Auer.2020,
    title = {{Improving Access to Scientific Literature with Knowledge Graphs}},
    author = {Sören Auer and Allard Oelen and Muhammad Haris and Markus Stocker and Jennifer D’Souza and Kheir Eddine Farfar and Lars Vogt and Manuel Prinz and Vitalis Wiens and Mohamad Yaser Jaradeh},
    volume = {44},
    number = {3},
    journal = {Bibliothek Forschung und Praxis},
    year = {2020},
    doi = {10.1515/bfp-2020-2042}
}

@article{Stocker.2023,
    title = {{FAIR Scientific Information with the Open Research Knowledge Graph}},
    author = {Markus Stocker and Allard Oelen and Mohamad Yaser Jaradeh and Muhammad Haris and Omar Arab Oghli and others},
    year = {2023},
    journal = {FAIR Connect},
    volume = {1},
    number = {1},
    publisher = {IOS Press},
    doi = {10.3233/FC-221513}
}

@incollection{Stroemert.2023,
  title={{Towards a Versatile Terminology Service for Empowering FAIR Research Data: Enabling Ontology Discovery, Design, Curation, and Utilization Across Scientific Communities}},
  author={Str{\"o}mert, Philip and Limbachia, Vatsal and Oladazimi, Pooya and Hunold, Johannes and Koepler, Oliver},
  booktitle={Knowledge Graphs: Semantics, Machine Learning, and Languages},
  pages={53--69},
  year={2023},
  publisher={IOS Press},
  doi = {10.3233/SSW230005}
}

@inproceedings{Karras.2021,
  title={{Researcher or Crowd Member? Why not both! The Open Research Knowledge Graph for Applying and Communicating CrowdRE Research}},
  author={Karras, Oliver and Groen, Eduard C and Khan, Javed Ali and Auer, S{\"o}ren},
  booktitle={2021 IEEE 29th International Requirements Engineering Conference Workshops (REW)},
  pages={320--327},
  year={2021},
  organization={IEEE},
  doi = {10.1109/REW53955.2021.00056}
}

@BOOK{Robinius.2020,
      author       = {Robinius, Martin and Markewitz, Peter and Lopion, Peter and
                      Kullmann, Felix and Heuser, Philipp and Syranidou, Chloi and
                      Cerniauskas, Simonas and Schöb, Thomas and Reuß, Markus
                      and Ryberg, Severin and Kotzur, Leander and Caglayan, Dilara
                      and Welder, Lara and Linßen, Jochen and Grube, Thomas and
                      Heinrichs, Heidi and Stenzel, Peter and Stolten, Detlef},
      title        = {{WEGE} {FÜR} {DIE} {ENERGIEWENDE} {K}osteneffiziente und
                      klimagerechte {T}ransformationsstrategien für das deutsche
                      {E}nergiesystem bis zum {J}ahr 2050},
      volume       = {499},
      publisher    = {Forschungszentrum Jülich GmbH Zentralbibliothek, Verlag},
      reportid     = {FZJ-2020-02537},
      isbn         = {978-3-95806-483-6},
      series       = {Schriften des Forschungszentrums Jülich Reihe Energie $\&$
                      Umwelt / Energy $\&$ Environment},
      pages        = {VIII, 141 S.},
      year         = {2020},
      url          = {https://juser.fz-juelich.de/record/877960},
}

@TechReport{Knublauch.2017,
  author = {Holger Knublauch and Dimitris Kontokostas},
  title  = {{Shapes Constraint Language (SHACL)}},
  url = {https://www.w3.org/TR/2017/REC-shacl-20170720/},
  year = {2017},
  type = {{W3C Recommendation}},
  institution = {W3C},
}

@misc{Goepfert.2022,
  doi = {10.48366/R150337},
  author = {Jan Göpfert and Oliver Karras},
  title = {{Comparison of Scenario Factsheets from the Open Energy Platform}},
  howpublished = {Open Research Knowledge Graph},
  year = {2022},
}

@misc{Kullmann.2021,
  doi = {10.48366/R153801},
  author = {Kullmann, Felix and Göpfert, Jan and Karras, Oliver and Kuckertz, Patrick and Auer, Sören and Stocker, Markus and Markewitz, Peter and Kotzur, Leander and Stolten, Detlef},
  title = {{Comparison of Studies on Germany's Energy Supply in 2050}},
  howpublished = {Open Research Knowledge Graph},
  year = {2021}
}

@article{Auer.2023,
title = {{The SciQA Scientific Question Answering Benchmark for Scholarly Knowledge}},
author = {Sören Auer and Dante A. C. Barone and Cassiano Bartz and Eduardo G. Cortes and Mohamad Yaser Jaradeh and Oliver Karras and Manolis Koubarakis and Dmitry Mouromtsev and Dmitrii Pliukhin and Daniil Radyush and Ivan Shilin and Markus Stocker and Eleni Tsalapati },
doi = {10.1038/s41598-023-33607-z},
year = {2023},
journal = {Nature Scientific Reports},
volume = {13},
number = {7240}
}

@inproceedings{Auer.2023a,
  title={{Organizing Scholarly Knowledge in the Open Research Knowledge Graph: An Open-Science Platform for FAIR Scholarly Knowledge}},
  author={Auer, S{\"o}ren and Stocker, Markus and Karras, Oliver and Oelen, Allard and D'Souza, Jennifer and Lorenz, Anna-Lena},
  booktitle={Proceedings of the Conference on Research Data Infrastructure},
  volume={1},
  year={2023},
  doi = {10.52825/cordi.v1i.272}
}

\end{document}